\documentclass[final,5p,times,twocolumn]{elsarticle}

\usepackage{epsfig}
\usepackage{epstopdf}

\usepackage{amssymb}

\journal{Journal of Physics and Chemistry of Solids}

\begin{document}

\begin{frontmatter}



\title{Effects of impurities and vortices on the low-energy spin excitations in high-T$_c$ materials}


\author[1]{Brian M. Andersen\corref{cor1}}\ead{bma@fys.ku.dk}
\author[2]{Markus Schmid}
\author[2,3]{Siegfried Graser}
\author[3]{P. J. Hirschfeld}
\author[2]{Arno P. Kampf}

\address[1]{Niels Bohr Institute, University of Copenhagen, Universitetsparken 5, DK-2100 Copenhagen, Denmark}
\address[2]{Theoretical Physics III, Center for Electronic Correlations and Magnetism, Institute of Physics, University of Augsburg,
D-86135 Augsburg, Germany}
\address[3]{Department of Physics, University of Florida, Gainesville, Florida 32611, USA}

\cortext[cor1]{Corresponding author.}

\begin{abstract}
We review a theoretical scenario for the origin of the spin-glass phase of underdoped cuprate materials. In particular it is shown how disorder in a correlated $d$-wave superconductor generates a magnetic phase by inducing local droplets of antiferromagnetic order which eventually merge and form a quasi-long range ordered state. When correlations
are sufficiently strong, disorder is unimportant for the generation of static magnetism but plays an additional role of pinning disordered stripe configurations. We calculate the spin excitations in a disordered spin-density wave phase, and show how disorder and/or applied magnetic fields lead to a slowing down of the dynamical spin fluctuations in agreement with neutron scattering and muon spin rotation ($\mu$SR) experiments.  
\end{abstract}

\begin{keyword}

High-T$_c$ superconductors, Disorder-induced magnetism, Field-induced magnetism, Neutron response, Stripes


\PACS 74.25.Ha \sep 74.72.-h \sep 75.40.Gb

\end{keyword}

\end{frontmatter}



\section{Introduction}
\label{}

From the phase diagram of the cuprate superconductors it is evident that as a function of doping, antiferromagnetic (AF) order gives way to $d$-wave superconductivity (dSC). This has led to the natural view that these two ordered phases generally compete in these materials. At an overall global scale this certainly seems to be the case. However, when including disorder the opposite can also occur: dSC can function as a catalyst for induced spin-density wave (SDW) order\cite{schmid}. This takes place because of the generation of low-energy impurity resonance states which are spin split by local magnetic ordering\cite{tsuchiura,wanglee,JWHarter}. A qualitatively similar effect happens in the vortex phase generated by an external applied magnetic field\cite{schmid}. The $d$-wave symmetry of the superconducting state is crucial for this cooperative effect to occur between SDW and dSC phases.

Experimentally, it is well known that disorder may cause slowing down and eventual freezing of spin fluctuations\cite{hirota,lake02,kimura03,savici}. For example, substitution of nonmagnetic Zn ions for Cu in near-optimally doped La$_{2-x}$Sr$_x$CuO$_4$ (LSCO) has been shown to shift spectral weight into the spin gap, and eventually, for enough Zn ($\sim 2\%$), generate elastic magnetic peaks in the neutron response\cite{kimura03}. In YBa$_2$(Cu$_{1-y}$Zn$_y$)$_3$O$_{6.97}$ with $y=2\%$ a similar in-gap Zn-induced spin mode was observed\cite{sidis96}. Upon increased temperature $T$ the elastic signal decreases and eventually vanishes near $T_c$ which is similar to an equivalent disorder signal in Zn-free LSCO\cite{lake02,kimura99}, These results of disorder-induced freezing of spin fluctuations are consistent with $\mu$SR data on underdoped cuprates\cite{mendels,bernhard,niedermayer,watanabe,panagopoulos}. Recent neutron scattering off detwinned YBa$_2$Cu$_3$O$_{6.6}$ (YBCO) with 2\% Zn found induced short-range magnetic order and a redistribution of spectral weight from the resonance peak to uniaxial incommensurate (IC) spin fluctuations at lower energies\cite{suchaneck}.

Applied magnetic fields introduce vortices into the system and cause much of the same 
phenomenology as described above for the disorder. Theoretically this can be ascribed to the fact that vortices in dSC also generate low-energy resonance states which may favorably split due to electronic correlations and generate local SDW order\cite{schmid,ogata,andersen00,zhu1,zhu2,ghosal,andersen03}.
For samples without static SDW order, experiments show that at low magnetic fields the vortices slow down the spin fluctuations and generate an in-gap mode\cite{lake01,chang09,andersen09}. For larger fields the ground state exhibits static local
SDW order which disappears above $T_c$. Enhancement of static order was
reported first in neutron diffraction experiments\cite{lake02,katano} on 
LSCO, with a correlation length
of several hundred \AA, and has been confirmed in other underdoped cuprates\cite{khaykovich02,khaykovich05,chang08,haug}. 

The freezing of spin fluctuations is also relevant even in the absence of substitutional disorder or external applied magnetic fields, as evidenced by the presence of a spin-glass phase in the underdoped regime. This seems to apply to both "clean" cuprates like YBCO where quasi-static SDW order is found\cite{stock06, sonier,stock08,hinkov}, and to intrinsically disordered materials like LSCO where the static spin correlations are long-range, and persist for a large doping range well into the dSC dome\cite{wakimoto,julien}. The size of the spin-glass phase in temperature and doping is clearly enhanced by disorder.  

Here, we review a picture that has emerged for the description of the induced magnetism in underdoped cuprate materials\cite{schmid,JWHarter,andersen07,andersenDresden,andersen08,andersen10}. We focus on the disorder- and field-induced SDW phase coexisting with dSC, and calculate the associated dynamical spin susceptibility.

\section{Model}

The basis for our model analysis is the BCS Hamiltonian for a
dSC with orbital coupling to an applied magnetic
field {\bf B}, to which we add site-centered disorder and a local
Hubbard repulsion treated in an unrestricted Hartree-Fock
approximation:
\begin{eqnarray}
H = - \sum_{ij\sigma} t_{ij} \: e^{{\rm i} \varphi_{ij}}\: c^{\dagger}_{i\sigma}
c^{}_{j\sigma} + \sum_{i\sigma} (V_i^{imp}-\mu)c^{\dagger}_{i\sigma} c^{}_{i\sigma}
\nonumber \\
+ \sum_{\langle ij \rangle} \left(\Delta_{ij} c^{\dagger}_{i\uparrow}
c^{\dagger}_{j\downarrow} + h.c.\right) + \frac{U}{2} \sum_i\left(\langle n_i \rangle n_i - \langle \sigma_i^z
\rangle \sigma_i^z \right).\label{Hamiltonian}
\end{eqnarray}
Here, $c^\dagger_{i\sigma}$ creates an electron on
site $i$ with spin $\sigma=\uparrow,\downarrow$ on a $N\times N$ square lattice.  The hopping matrix
elements between nearest and next-nearest neighbor sites are denoted
by $t_{ij} = t$ and $t_{ij} = t'$, respectively. The magnetic field is included through standard Peierls phase factors $\varphi_{ij}=(\pi/\Phi_0)
\int^{{\bf r}_i}_{{\bf r}_j} {\bf A}({\bf r})\cdot{\rm d}{\bf r}$, where
$\Phi_0 = hc/(2e)$ and ${\bf
A(r)} = B (0, x)$ is the vector potential in
Landau gauge. The chemical potential $\mu$ is adjusted to fix the
electron density $n = \frac{1}{N^2} \sum_i \langle n_i \rangle = 1 - x$,
where $x$ is the hole concentration. The $d$-wave pairing
amplitude $\Delta_{ij}$ is determined by the
strength of an attractive nearest-neighbor interaction $V_d$. The
impurity potential $V_i^{imp}$ arises from a concentration $n_{imp}$ of nonmagnetic 
pointlike scatterers at random positions. All fields, $\Delta_{ij}$,
the local charge density $\langle n_i \rangle$, and the local
magnetization $\langle \sigma_i^z
\rangle$ are calculated self-consistently from the Bogoliubov-de Gennes (BdG) equations\cite{JWHarter,ghosal}. In the following we fix parameters to 
$t'=-0.4t$, $V_d=1.34t$. 

The model (\ref{Hamiltonian}) has been used
extensively to study the competition between dSC and SDW phases\cite{martin,andersen05}, the origin of
field-induced magnetization\cite{schmid,andersen00,zhu1,zhu2,ichioka}, and moment formation around nonmagnetic impurities\cite{JWHarter,andersen07,andersen08}. In the 
case of many impurities, Eq.(\ref{Hamiltonian}) was used to explain the origin of static disorder-induced magnetic droplet phases\cite{andersen07,andersen08,alvarez,atkinson}, and to study how these may form a quasi-long range ordered SDW phase. The underlying idea is that when magnetic droplets begin to 
overlap, either when  the magnetic correlation length grows at low doping, or when the concentration of scatterers is increased, 
the effective interaction between droplets allows them to align their staggered patterns coherently\cite{andersen07,shender}.
More recently Eq.(\ref{Hamiltonian}) was used to obtain semi-quantitative description of the $T$-dependence of the elastic neutron response in LSCO in an applied magnetic field\cite{schmid,lake02}. Finally, Eq.(\ref{Hamiltonian}) has been used to explain various transport measurements\cite{andersen08,weichen} such as, for example, the experimental observation of a non-universal low-$T$ limit of the thermal conductivity $\kappa(T)$ \cite{sutherland,sun} in terms of a disorder-induced SDW phase\cite{andersen08}. The SDW droplet phase has an enhanced scattering rate while maintaining the same low energy density of states (DOS) as the homogeneous case, thus breaking the cancellation between the residual quasiparticle DOS and relaxation rate
which gave rise to the universal $\kappa(T)$ in the first place\cite{graf,durstlee}. Thus, for underdoped cuprates it is not $\kappa(T)$ but rather the (spatially averaged) low-energy DOS which is universal, in agreement with STM experiments\cite{lang}. Theoretically a universal low-energy DOS can be traced to suppressed charge modulations caused by the electronic Coulomb repulsion\cite{andersenDresden,andersen08,garg}.

\section{Results and Discussion}

The Hamiltonian (\ref{Hamiltonian}) supports both a correlation- and disorder-induced SDW phase. Specifically, in the clean case above a critical repulsion $U_{c2}$ a global stripe phase is the favorable state, and disorder acts mainly to scramble the stripes. Below $U_{c2}$ the ground state is a homogeneous dSC but nonmagnetic disorder or vortices may locally induce SDW order if $U>U_{c1}$ where $U_{c1}$ is another critical interaction strength\cite{schmid,JWHarter}. In this low-$U$ regime, both disorder and the $d$-wave symmetry of the pairing are crucial for generating static magnetism. 

\begin{figure}[h]
\begin{minipage}{.49\columnwidth}
\includegraphics[clip=true,height=0.8\columnwidth,width=0.98\columnwidth]{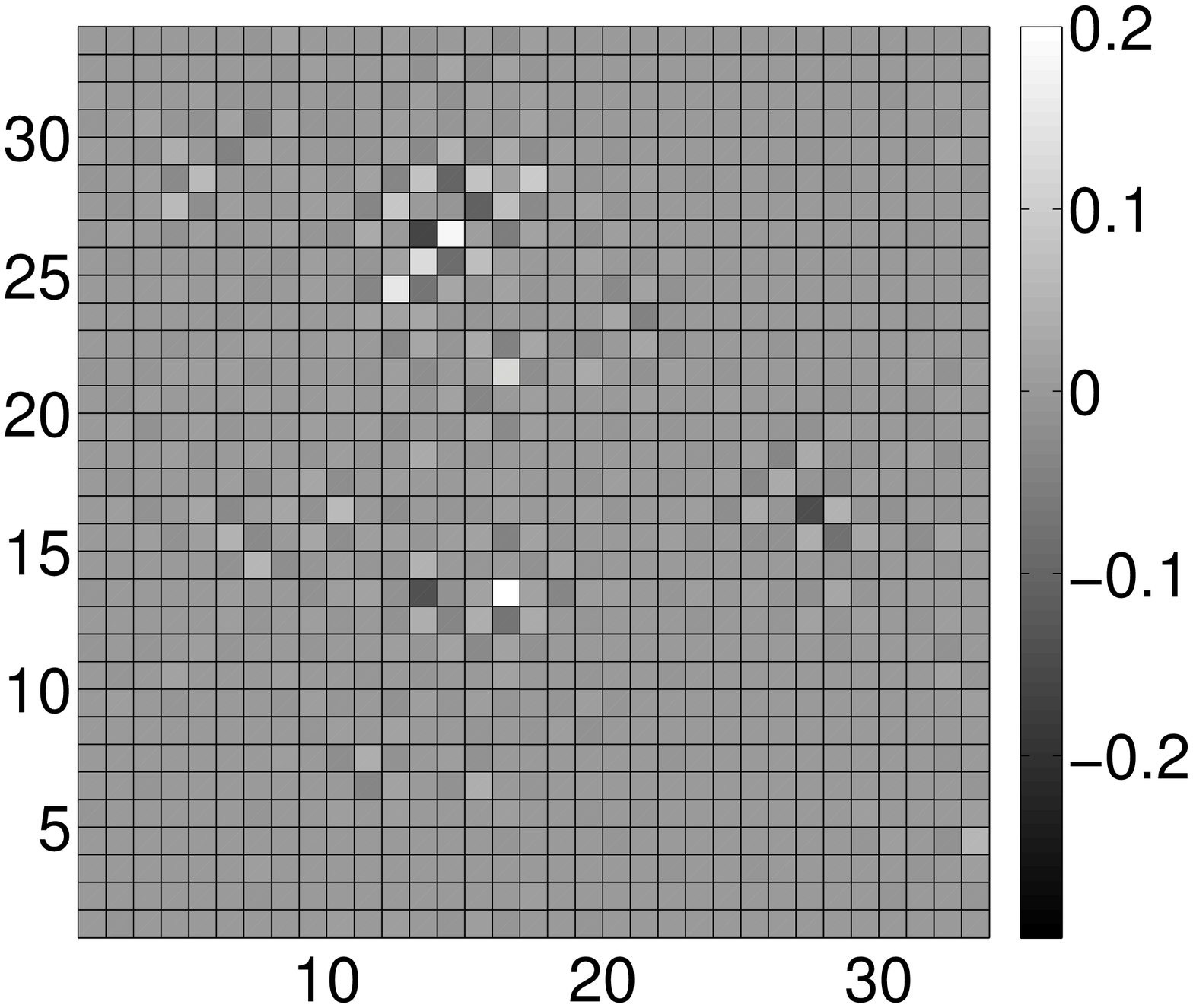}
\put(-126,87){{\bf a}}
\end{minipage}
\begin{minipage}{.49\columnwidth}
\includegraphics[clip=true,height=0.8\columnwidth,width=0.98\columnwidth]{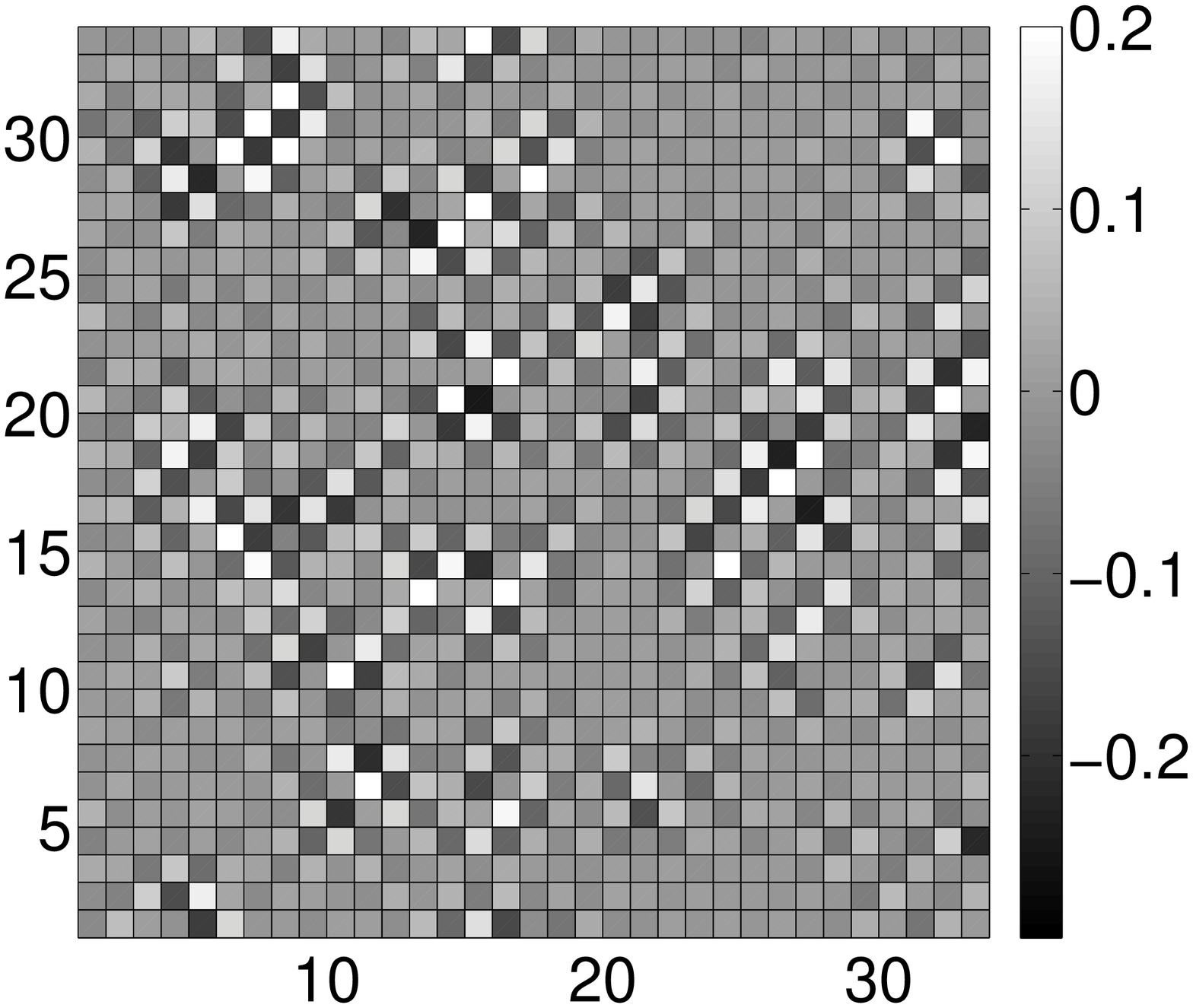}
\put(-126,87){{\bf b}}
\end{minipage}
\\
\begin{minipage}{.49\columnwidth}
\includegraphics[clip=true,height=0.8\columnwidth,width=0.98\columnwidth]{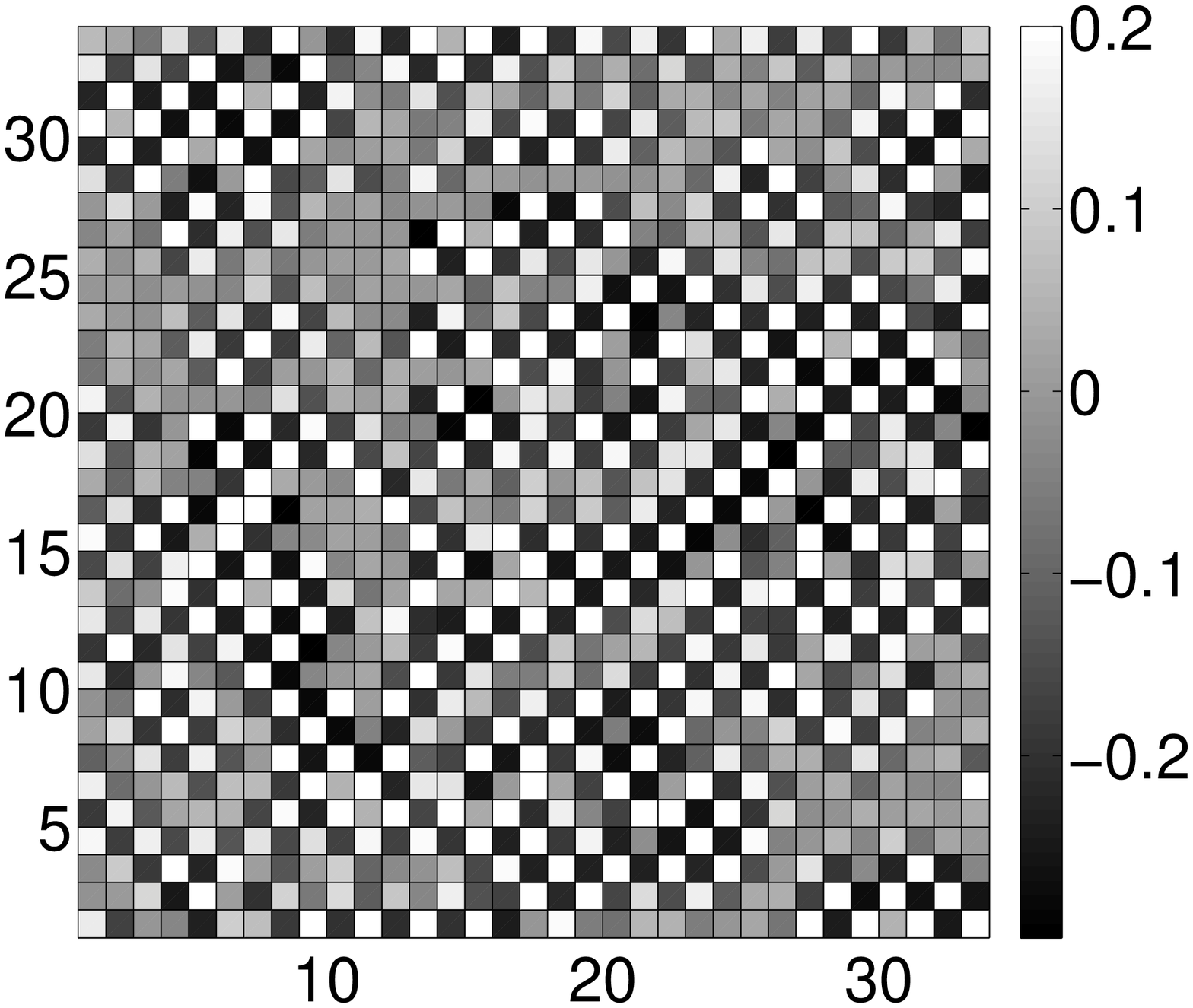}
\put(-126,87){{\bf c}}
\end{minipage}
\begin{minipage}{.49\columnwidth}
\includegraphics[clip=true,height=0.8\columnwidth,width=0.98\columnwidth]{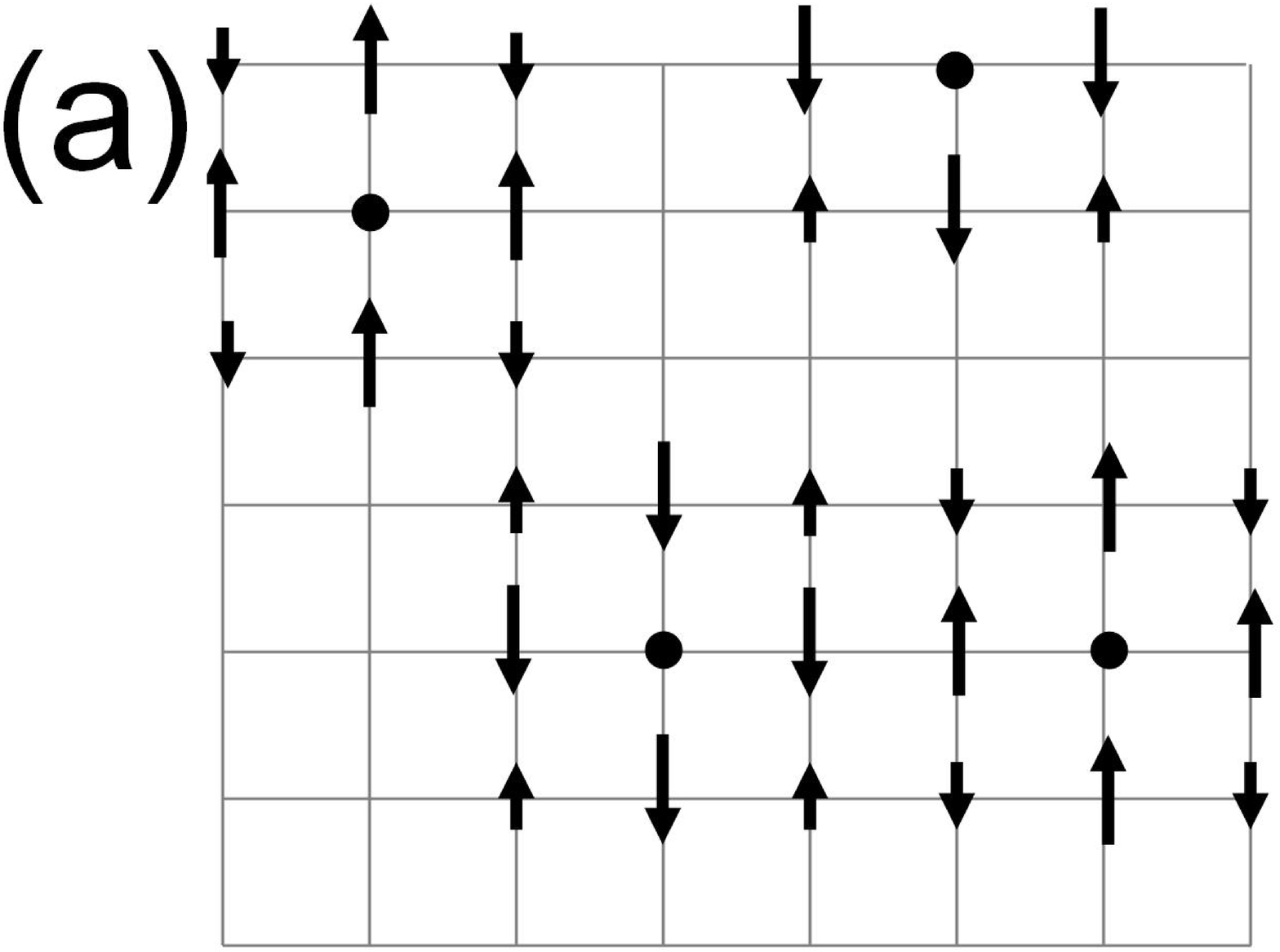}
\put(-126,87){{\bf d}}
\end{minipage}
\caption{(a-c) Disorder-induced real-space magnetization from a single random weak impurity configuration with $B=0$, $V^{imp}=3t$, $T=0.025t$, $n_{imp}=x=7.5\%$ and $U=2.8t$ (a), $U=3.2t$ (b), and $U=3.6t$ (c). (d) Schematic illustration of how a single N{\'e}el phase gets stabilized by disorder.} \label{fig1}
\end{figure}

In the absence of a magnetic field, Fig. \ref{fig1}(a-c) show the magnetization 
in real-space arising from a weak disorder potential from the dopant ions\cite{andersen07}. 
Figure \ref{fig1}(a) shows that not all impurities in the
correlated system need to "magnetize" for a given $U$; in
the disordered system, the effective criterion to drive the
impurity through the local magnetic phase transition is different
for each impurity. Increasing the repulsion $U$ then increases the
concentration of impurities which induce a local magnetization
droplet, as shown in Fig. \ref{fig1}(b) and \ref{fig1}(c). With further
increase of $U$, the system evolves from a state with dilute
non-overlapping AF droplets to connected spin textures.  A similar process takes place in 1D where Shender
and Kivelson\cite{shender} pointed out that the interactions
between impurities in a quantum spin chain are non-frustrating: if
an impurity creates a local AF droplet, a second one can always
orient itself to avoid losing exchange energy. In 2D  this continues to apply for spin models with nearest neighbor
exchange, but may break down in the presence of mobile charges.

The present mean field treatment of the Hubbard interaction does not
capture the band narrowing due to correlations near
half-filling leading to the Mott transition. However, it is expected
that underdoped systems are characterized by larger effective
interactions.  Within the present scenario for dirty cuprates, the $x$ dependence of the
spin order is therefore described qualitatively by the sequence
\ref{fig1}(c) $\rightarrow$ \ref{fig1}(b)
$\rightarrow$ \ref{fig1}(a), until it disappears completely
at effective $U$'s below $U_{c1}$ near optimal doping. Increasing
$x$ should also be accompanied by a weakening of the dopant disorder potential
 due to enhanced screening. Within our model, increasing
$U$ or $V^{imp}$ leads to qualitatively similar results, and we
cannot determine from this approach which effect is dominant in real
systems. Note from Fig. \ref{fig1} that AF droplets are
induced for $U\gtrsim 2.4t$, a substantially reduced critical value
compared to the 1-impurity case where $U_{c1}=3.25t$. This is
because the Hubbard correlations induce charge redistributions which
alter the effective local chemical potential, such that the
criterion for magnetization of each impurity site depends on its
local disorder environment. Some regions containing impurities have
charge densities closer to the phase boundary for AF order, thus
enhancing local moment formation relative to the single impurity
case. In the limit of large $U$,  the  magnetic order becomes
qualitatively similar to that arising in a stripe state with
quenched disorder\cite{andersen10,kaul,robertson,delmaestro}.

\begin{figure}[ht]
\includegraphics[width=8cm]{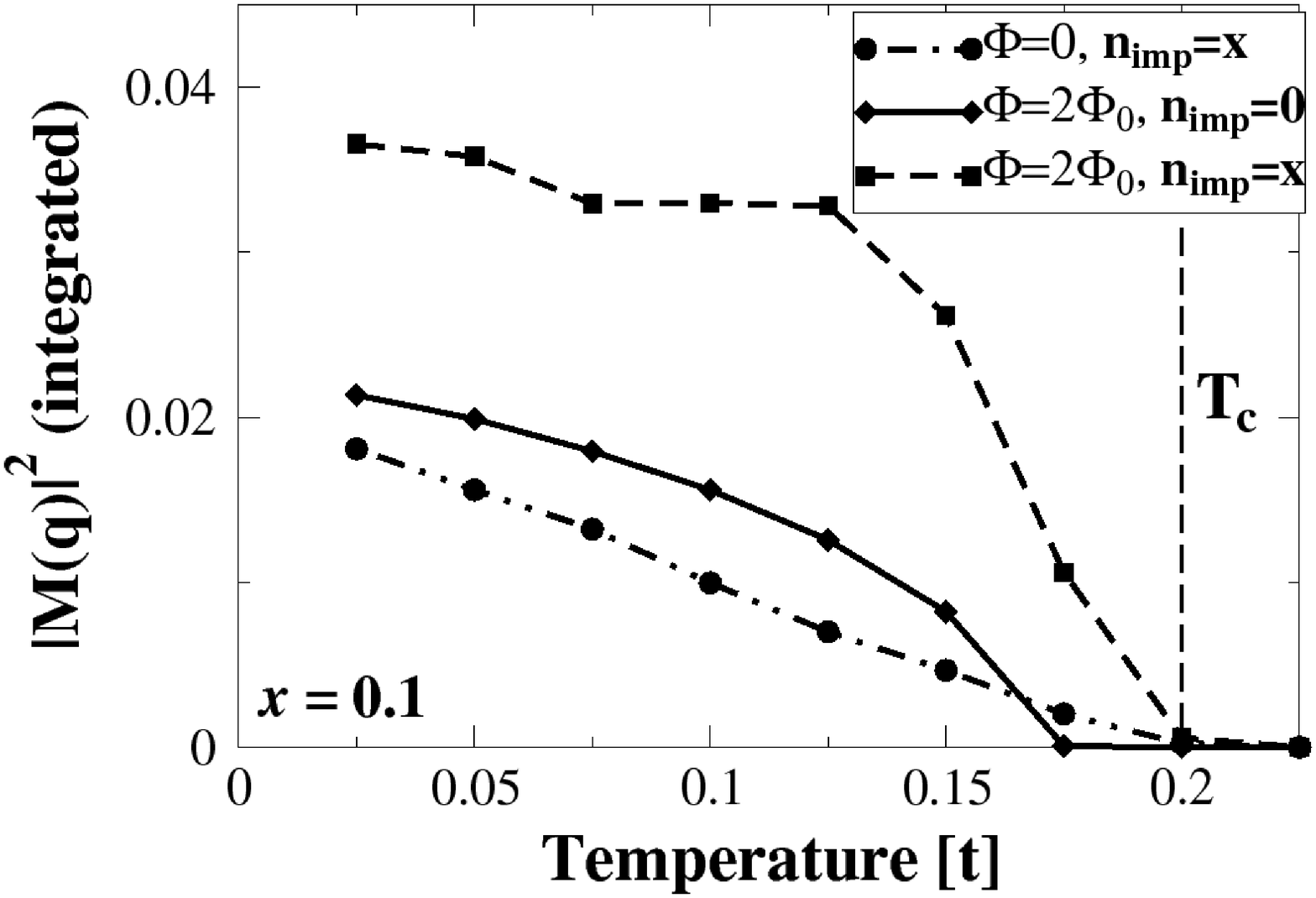}
\put(-230,139){{\bf a}}
\\
\\
\begin{minipage}{.49\columnwidth}
\includegraphics[clip=true,height=0.8\columnwidth,width=0.98\columnwidth]{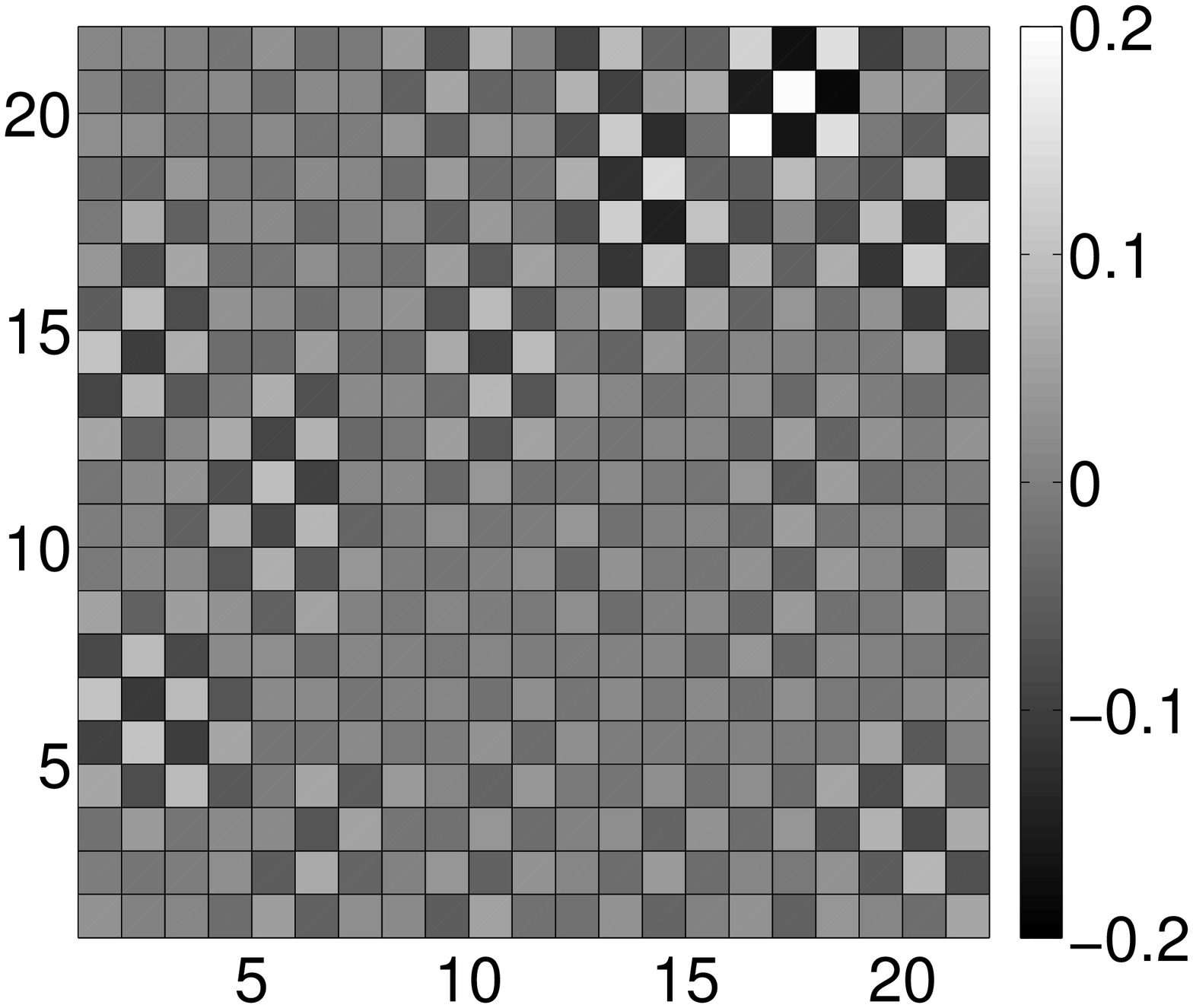}
\put(-126,87){{\bf b}}
\end{minipage}
\begin{minipage}{.49\columnwidth}
\includegraphics[clip=true,height=0.8\columnwidth,width=0.98\columnwidth]{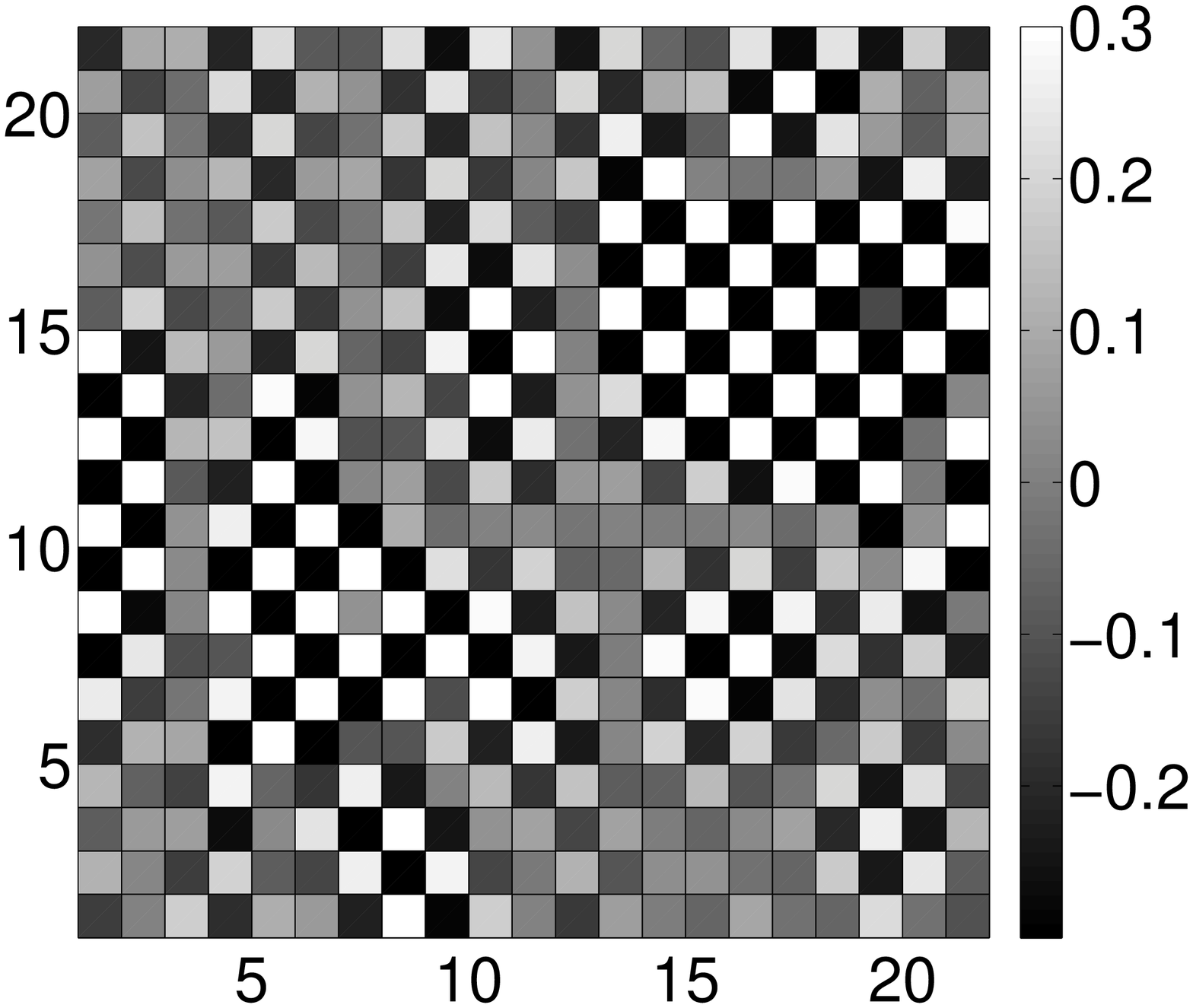}
\put(-126,87){{\bf c}}
\end{minipage}
\caption{(a) $T$-dependence of the peak intensity of the magnetic structure factor integrated around $(\pi,\pi)$ in zero field and finite field with the density 
of impurities $n_{imp} = x = 10\%$ (dash-dotted and dashed curves, respectively). For the data with $\Phi=2\Phi_0$ and $n_{imp} = x$, the zero-field data were subtracted. For comparison, the 
structure factor in a clean system is also included for the same magnetic-field strength (solid curve). 
$|M(q)|^2$ (integrated) translates directly to the ordered spin moment squared in units of $\mu_B$ per Cu$^{2+}$ ion. (b,c) Real-space magnetization at $T=0.025t$ for a system with 10\% weak impurities modeling a dopant potential (b) and with the addition of two vortices (c). In (c) the vortices are located at sites $(7,6)$ and $(15,17)$. For all results in this figure $U=2.9t$.} \label{fig2}
\end{figure}

In the presence of a magnetic field, vortices may lead to an enhancement of the SDW order. This is illustrated in Fig. \ref{fig2} which shows the magnetization for a random disorder potential with $B=0$ [\ref{fig2}(b)] and $B\neq 0$ [\ref{fig2}(c)]. In Fig. \ref{fig2}(c) two flux quanta penetrate the system corresponding to a large magnetic field of $B=47 T$. Focussing on the $T$-dependence, Fig. \ref{fig2}(a) shows the structure factor $|M(q)|^2$ integrated near $q=(\pi,\pi)$ as a function of $T$\cite{schmid}. The main result of Fig. \ref{fig2}(a) is seen by the dashed and dot-dashed lines displaying  $|M(q)|^2$ including disorder with and without an applied magnetic field, respectively. In both cases the magnetic order sets in at $T_c$ below which the $T$-dependence is in remarkable agreement with neutron diffraction data on underdoped LSCO\cite{lake02}. The origin of the qualitatively different curvature of the $T$-dependence of $|M(q)|^2$ in the presence of a magnetic field can be traced to a larger concentration of anti-phase domain walls when $B=0$\cite{schmid}.

We now turn to a discussion of the transverse dynamical susceptibility $\chi_0^{xx}(\vec{r}_{i},\vec{r}_{j},\omega)=-i\int_{0}^{\infty}dt\, e^{i\omega t}\left\langle \left[\sigma_{i}^{x}(t),\sigma_{j}^{x}(0)\right]\right\rangle$ which can be expressed in terms of the BdG eigenvalues $E_n$ and eigenvectors $u_n, v_n$
\begin{eqnarray}\nonumber
\chi_0^{xx}(\vec{r}_{i},\vec{r}_{j},\omega) & = & \sum_{m,n} f(u,v)\frac{f(E_{m})+f(E_{n})-1}{\omega+E_{m}+E_{n}+i\Gamma}\\
 & + & \sum_{m,n} g(u,v)\frac{f(E_{m})+f(E_{n})-1}{\omega-E_{m}-E_{n}+i\Gamma},
\end{eqnarray}
where $f(u,v) = u_{m,i}^{*}v_{n,i}^{*}\left(u_{m,j}v_{n,j}-u_{n,j}v_{m,j}\right)$, and
$g(u,v) = v_{m,i}u_{n,i}\left(u_{m,j}^{*}v_{n,j}^{*}-u_{n,j}^{*}v_{m,j}^{*}\right)$.
Including the electronic interactions within RPA we find for the full susceptibility
\begin{equation}
\chi^{xx}(\vec{r}_i,\vec{r}_j,\omega)\!=\!\sum_{\vec{r}_l}\left[1-U\chi^{xx}_{0}(\omega)\right]_{\vec{r}_i,\vec{r}_l}^{-1}\chi^{xx}_{0}(\vec{r}_l,\vec{r}_j,\omega).
\end{equation}
Fourier transforming with respect to the relative coordinate $\vec{r}=\vec{r}_i-\vec{r}_j$ defines the spatially resolved momentum-dependent
susceptibility $\chi(\vec{q},\vec{R},\omega)=\sum_{\vec{r}}e^{i\vec{q}\cdot\vec{r}}\chi(\vec{R},\vec{r},\omega)$. Averaging over the center of mass coordinate $\vec{R}=(\vec{r}_i+\vec{r}_j)/2$, this expression gives the susceptibility $\chi(\vec{q},\omega)$ relevant for comparison with neutron scattering measurements. We have checked that at half-filling in the clean AF state without pairing, $\chi(\vec{q},\omega)$ is characterized by spin-wave excitations on a cone centered at $(\pi,\pi)$ as expected. 

\begin{figure}[h]
\begin{minipage}{.49\columnwidth}
\includegraphics[clip=true,height=0.8\columnwidth,width=0.98\columnwidth]{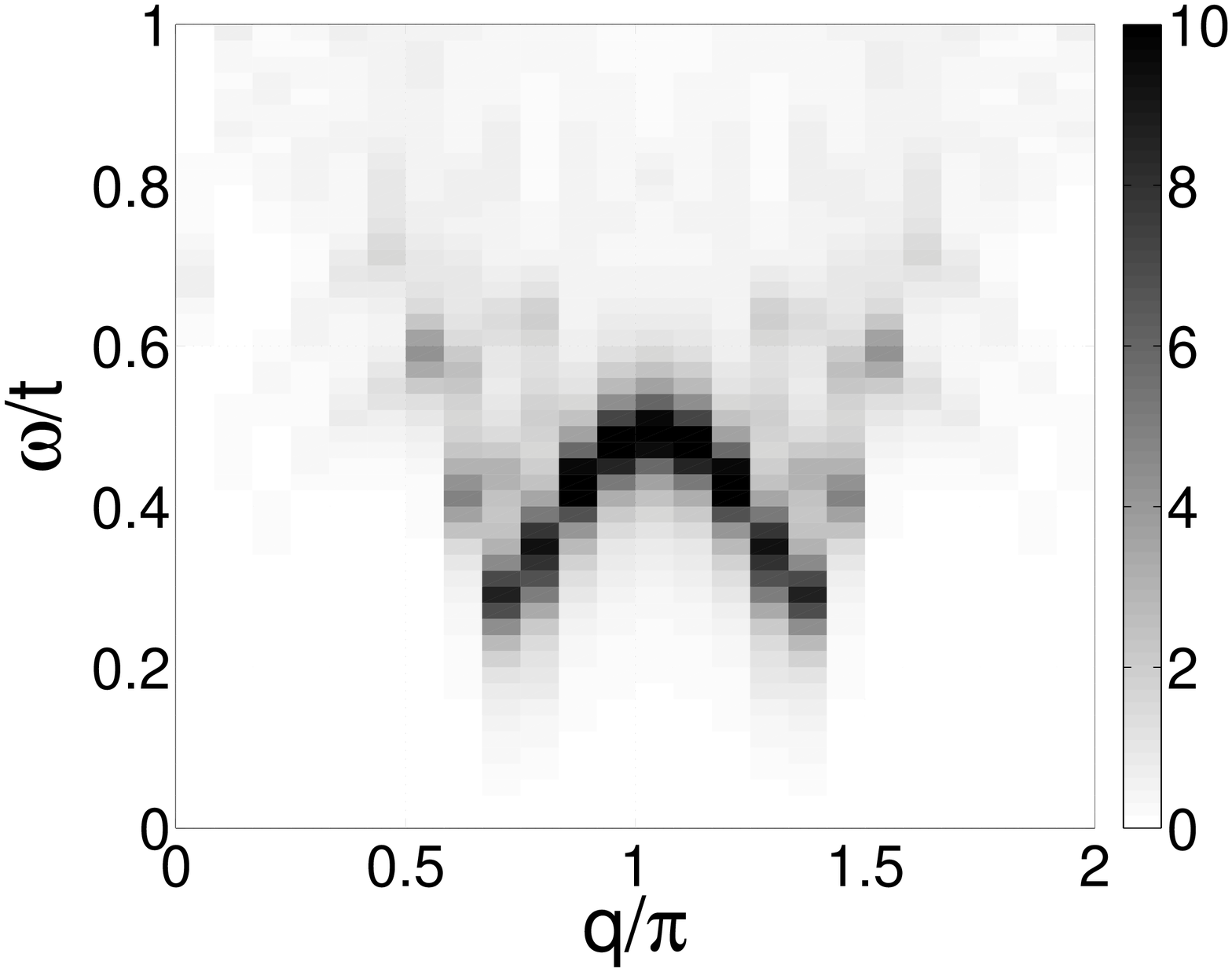}
\put(-126,87){{\bf a}}
\end{minipage}
\begin{minipage}{.49\columnwidth}
\includegraphics[clip=true,height=0.8\columnwidth,width=0.98\columnwidth]{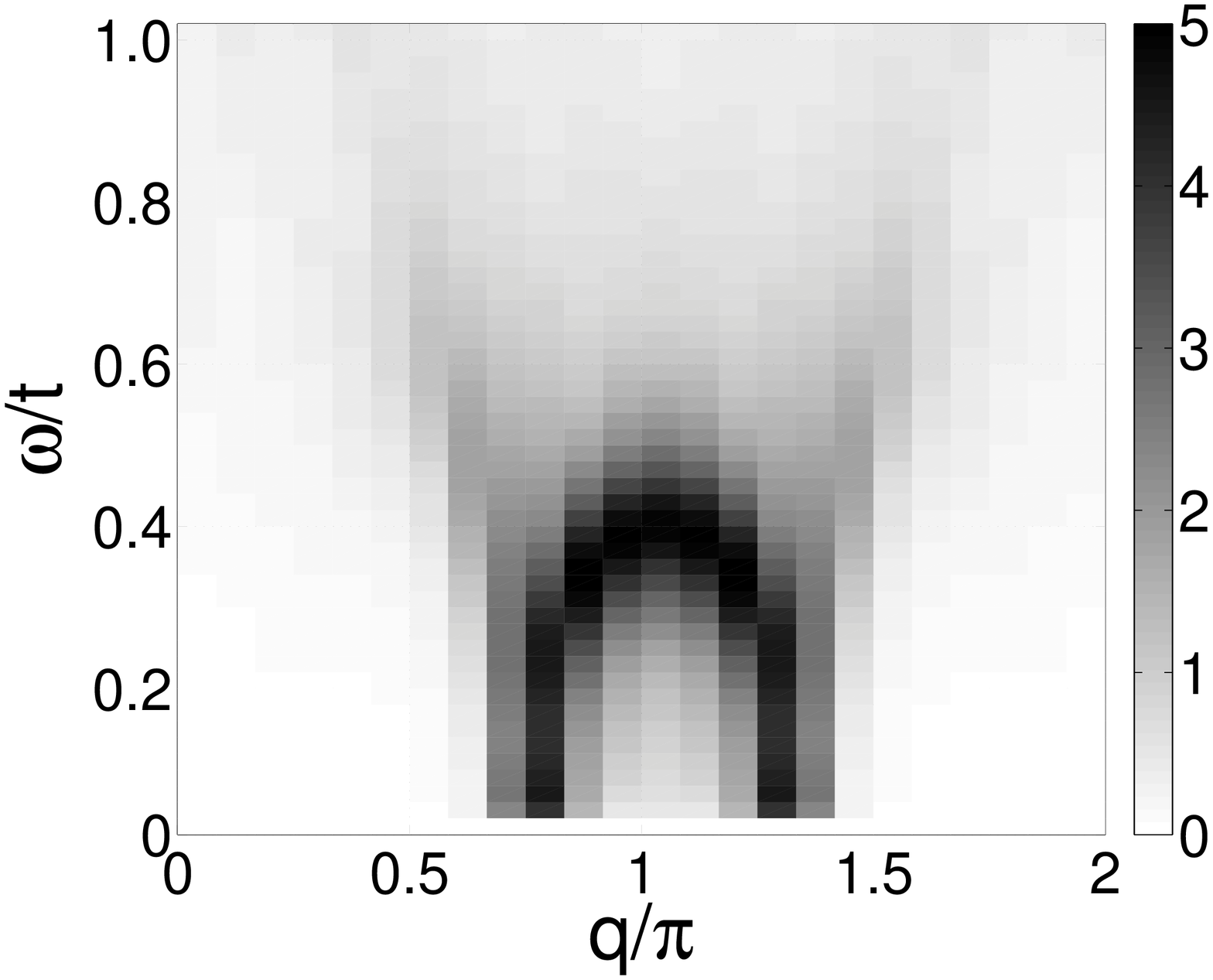}
\put(-126,87){{\bf b}}
\end{minipage}
\\
\begin{minipage}{.49\columnwidth}
\includegraphics[clip=true,height=0.8\columnwidth,width=0.98\columnwidth]{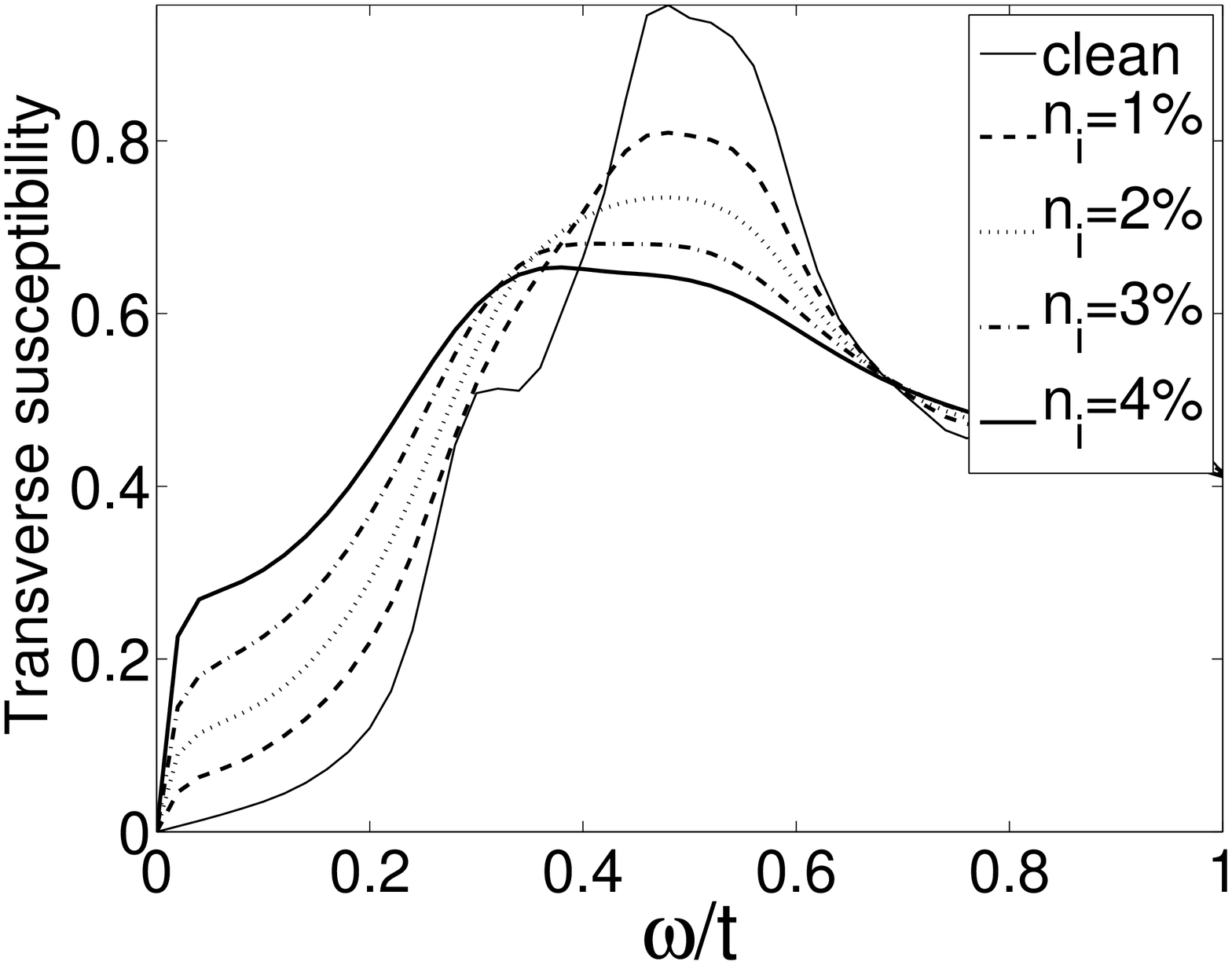}
\put(-126,87){{\bf c}}
\end{minipage}
\begin{minipage}{.49\columnwidth}
\includegraphics[clip=true,height=0.8\columnwidth,width=0.98\columnwidth]{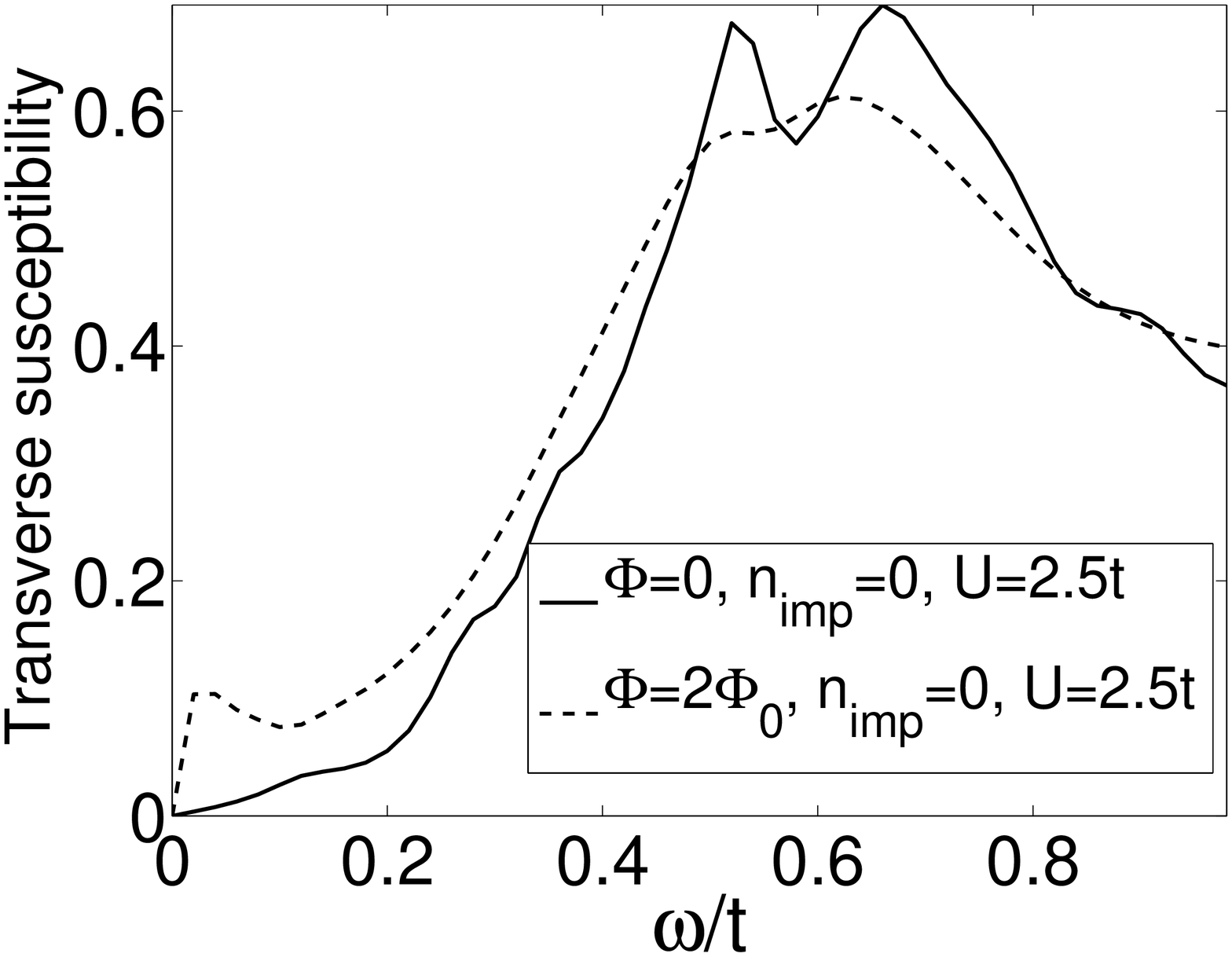}
\put(-126,87){{\bf d}}
\end{minipage}
\caption{(a,b) Spin susceptibility $\chi(q,\pi,\omega)$ for a clean (dirty) dSC phase without (with) disorder using $U=2.6t$, $B=0$, $\Gamma=0.04t$, and in (b) $n_{imp}=3\%$ strong scatterers with $V^{imp}=100t$. (c) Local susceptibility $\chi(\omega)$ for the same parameters as in (b) except for different disorder concentrations of strong scatterers. In (b,c) we have averaged over 5 different impurity configurations. (d) comparison of $\chi(\omega)$ in the clean dSC phase with (without) an applied magnetic field shown by dashed (solid) line.} \label{fig3}
\end{figure}

Although we are restricted to a mean-field/RPA approach, 
a strength of the present calculation is that we can include both spin and charge degrees of freedom 
within an unrestricted method which is capable of describing realistic inhomogeneous situations, 
and thus naturally includes both the response from regions dominated by the dSC condensate or local SDW order. For a homogeneous dSC our model reduces to a system of Bogoliubov quasiparticles whose magnetic response displayed in Fig. \ref{fig3}(a) depends crucially on the presence of a dSC gap. In the remainder, we investigate the role of spatially inhomogeneous local moments induced in this quasiparticle system. Figure  \ref{fig3}(b) shows how the magnetic droplet phase freezes the low-energy spin fluctuations by removing the spin-gap of the hour glass spectrum\cite{andersen10}.  This is also seen from the $q$-summed local susceptibility  $\chi(\omega)= \sum_{\vec{q}}\chi(\vec{q},\omega)$ plotted in Fig. \ref{fig3}(c) for different concentrations of the nonmagnetic disorder. In Fig. \ref{fig3}(b) and  \ref{fig3}(c) we have not included the weaker disorder potential arising from the dopants which will lead to additional freezing, but only included disorder from a few percent strong scatterers relevant for the modeling of e.g. Zn substitutional disorder. In Figs. \ref{fig3}(a,b,c) the spectrum is $C_4$ symmetric both in the static and dynamical response. In Ref. \cite{andersen10} is was shown how the stripe phase breaks this $C_4$ symmetry in the inelastic response even in the presence of substantial disorder concentrations.

We end by discussing the dynamical spin susceptibility when $B \neq 0$. Since the vortices can also nucleate 
local SDW order, we expect a similar slowing down of the fluctuations as for the point-like disorder. That this is the case can be seen from 
Fig. \ref{fig3}(d) which compares $\chi(\omega)$ for the clean dSC with and without an applied magnetic field\cite{footnote}. Clearly, the same 
filling-in of the low-energy weight takes place because of the vortices.

\section{Conclusions}

We have reviewed a theoretical scenario for the understanding of disorder- and magnetic field-induced SDW phases in underdoped cuprate superconductors. 
In the weak coupling limit, nonmagnetic impurities nucleate static SDW order. The $d$-wave nature of the pairing state is 
crucial for generation of local magnetism and works as a catalyst for the SDW phase. In the strong coupling limit disorder 
mainly acts to scramble correlation-generated stripe order. We demonstrated explicitly how both disorder and magnetic fields slow down the spin excitations
as observed in experiments. 


\section*{Acknowledgements}

B.M.A. acknowledges support from The Danish Council for Independent Research $|$ Natural Sciences. M.S., S.G., and A.P.K. 
acknowledge support from the DFG through TRR 80.
P.J.H. received partial support from the U.S.  Dept. of Energy  under grant
DE-FG02-05ER46236.




\begin{thebibliography}{00}
\bibitem{footnote} Note that in Fig. \ref{fig3}(d) we have a slightly larger $\Delta_{ij}\simeq 0.13$, and U=2.15 for the results shown there.
\end{thebibliography}
\end{document}